\documentstyle[preprint,revtex]{aps}
\begin {document}
\draft
\preprint{UCI TR 92-28 /Uppsala U. PT16 1992}
\begin{title}
Hadronic Instabilities in Very Intense Magnetic Fields
\end{title}
\author{Myron Bander\footnotemark\ }
\begin{instit}
Department of Physics, University of California, Irvine, California
92717, USA
\end{instit}
\author{H. R. Rubinstein\footnotemark\ }
\addtocounter{footnote}{-1}\footnotetext{e-mail:
mbander@ucivmsa.bitnet;
mbander@funth.ps.uci.edu}\addtocounter{footnote}{1}%
\footnotetext{e-mail:
rub@vand.physto.se}
\begin{instit}
Department of Radiation Sciences, University of Uppsala, Uppsala,
Sweden
\end{instit}
\centerline{}
\centerline {Submitted to the 26th International Conference on High
Energy Physics}
\centerline{Dallas, Texas, August, 1992}
\begin{abstract}
Composite hadronic states exhibit interesting properties in the
presence of very intense magnetic fields, such as those conjectured
to exist in the vicinity of certain astrophysical objects. We discuss
three scenarios. (i) The presence of vector particles with anomalous
magnetic moment couplings to scalar particles,
induces an instability of the vacuum. (ii) A delicate interplay
between the anomalous magnetic moments of the  proton and neutron
makes, in magnetic fields $B\ge 2\times 10^{14}$ T, the neutron
stable and for fields $B\ge 5\times 10^{14}$ T the proton becomes
unstable to a decay into a neutron via $\beta$ emission. (iii) In
the unbroken chiral $\sigma$ model magnetic fields would be screened
out as in a superconductor. It is the explicit breaking of chiral
invariance that restores standard electrodynamics. Astrophysical
consequences of all these phenomena are discussed.
\end{abstract}
\newpage
\narrowtext

\section{Introduction}
Very intense magnetic fields have been conjectured to exist in
connection with several astrophysical phenomena. Supernova may
contain fields up to $10^{10}$ T \cite{supernova}; a recent model for
extrgalactic gamma ray bursts \cite{Piran} involves fields of
$10^{13}$ T and fields of strengths greater than $10^{14}$ T are
associated with cosmic strings \cite{Berezinsky}. The latter are due
to \cite{Witten} currents $I=10^{20}$ A in strings whose
thickness is $1/M_W$. The presence of fields with electromagnetic
couplings of non-electromagnetic origin induces instabilities in
such large fields. The electro-weak theory itself, due to anomalous
magnetic moments of the charged gauge fields, produces vacuum
instabilities in the presence of fields of $10^{20}$ T
\cite{Ambjorn}. In this report we wish to summarize effects due to
ordinary hadronic physics which will manifest themselves in much
lower fields, $B=10^{14}$ T. We will discuss three such
phenomena: (i) the presence of magnetic dipole transitions between
hadronic resonances induces a  vacuum instability \cite{BanRub1};
(ii)~in fields of such magnitude, due to a subtle interplay of the
anomalous magnetic moments of the proton and neutron, the former
becomes heavier than the latter and decays into it by positron
emission \cite{BanRub2};  (iii) chiral symmetry breaking screens
magnetic fields larger than $10^{14}$ T \cite{BanRub3}. We shall
discuss each of these topics in turn and some possible consequences
at the end.

\section{Instabilities due to Magnetic Dipole Transitions}
Among the low lying hadronic states are scalar and vector mesons
with opposite parities and charge conjugations coupled to each
other by magnetic dipole transitions. We will point out that this
will lead to an instability in a large magnetic field.
The simplest model to exhibit this effect
contains a neutral scalar field $s(x)$ with mass $\mu$ and a
vector field $v^{\mu}(x)$ with mass $M$ and with parity and charge
conjugation opposite to that of $s$. An magnetic dipole transition
couples the electromagnetic field to $s$ and $v$. The Lagrangian for
such a model is
\begin{eqnarray}
{\cal L}=&&{1\over 2}\partial_{\mu}s\partial^{\mu}s-{1\over 2}\mu^2s^2-
{1\over4}
(\partial_{\mu}v_{\nu}-
\partial_{\nu}v_{\mu})(\partial^{\mu}v^{\nu}-\partial^{\nu}v^{\mu})
+{1\over 2}M^2v_{\mu}v^{\mu}
\nonumber\\
&&-{1\over 2}{e\over \Lambda}s(\partial_{\mu}v_{\nu}-
\partial_{\nu}v_{\mu})F^{\mu\nu}\, ;\label{lag}
\end{eqnarray}
$F^{\mu\nu}$ is the electromagnetic field strength tensor and
the last term is the aforementioned magnetic dipole coupling. $\Lambda$
has the dimensions of a mass and its magnitude provides the strength
of this coupling. For $F^{\mu\nu}$ constant
we can easily find the
eigenmodes of this Lagrangian.
The two transverse states of the vector meson remain uncoupled while
the longitudinal state mixes with the scalar particle yielding two
modes whose dispersion relation is obtained from a solution of
\begin{equation}
(p^2-\mu^2)(p^2-M^2)+{{e^2}\over
\Lambda^2}p_{\lambda}p_{\eta}F^{\lambda\nu}{F^{\eta}}_{\nu}=0\, .
\label{disp}
\end{equation}
 For a constant magnetic field pointing in the z direction
Eq.\ (\ref{disp}) takes the form
\begin{equation}
(p^2-\mu^2)(p^2-M^2)-{{e^2}\over \Lambda^2}{\bf {p_{\perp}}}^2H^2=0\, ;
\label{disp2}
\end{equation}
${\bf p_{\perp}}^2={p_x}^2+{p_y}^2$.
The energies satisfy
\begin{equation}
p_0^2={\bf p}^2\pm \left[\left ({{M^2-\mu^2}\over 2}\right
)^2+{{e^2}\over \Lambda^2}{\bf
{p_{\perp}}}^2H^2\right]^{1\over 2}\, .
\end{equation}
For sufficiently large magnetic fields the lower solution becomes
negative indicating an instability. This will occur whenever
\begin{equation}
H\ge H_c={{(M+\mu)\Lambda}\over {|e|}}\, .\label{hcrit}
\end{equation}

The instability will manifest itself in that fields $H\ge H_c$ will
create pairs of the $s$ and $v$ particles. $\Gamma$, the decay rate
per unit time per unit volume of a
``vacuum'' with such a large field may be calculated by standard
methods \cite{IZ}
\begin{equation}
\Gamma=Re\int {{d^4p}\over {(2\pi)^4}}\ln\left [
(p^2-\mu^2)(p^2-M^2)-{{e^2}\over \Lambda^2}{\bf
{p_{\perp}}}^2H^2\right ]\, .
\end{equation}
It is only for $H\ge H_c$ that the above integral develops a real
part
\begin{equation}
\Gamma={1\over {96\pi}}{{eH}\over\Lambda} \left\{\left({{eH}\over
\Lambda}\right)^2-2\left(M^2+\mu^2\right)+\left[{{\left(M^2-\mu^2\right)\Lambda}
   \over{eH}}\right]^2\right\}^{3\over2}\, .
\end{equation}

The masses involve are of the order of a few hundred MeV; we
likewise expect $\Lambda$ to be of the same order and thus the
critical fields will be in the range of $10^{15}$ T. Zeeman
splittings in such fields will be of the order of tens of MeV's
and we expect the effective Lagrangian of Eq.~(\ref{lag}) to be
valid.

\section{Proton $\beta$ Decay in Large Fields}
In this section we shall show how
a delicate interplay between the anomalous magnetic moments of the
proton and neutron makes, in magnetic fields $B\ge 2\times 10^{14}$ T,
the neutron stable and for fields $B\ge 5\times
10^{14}$ T the proton becomes unstable to a decay into a neutron via
$\beta$ emission.

The spectrum of Dirac particles with anomalous magnetic moments,
placed in uniform external magnetic fields can be obtained in a
straightforward manner; the previous statement is true as long as
the energy shifts due to the anomalous part of the moments are
smaller than the masses of the particles. For the field strengths
we shall study, this will be true for the proton and for the
neutron. This treatment has to be modified for the electron, as
even in these fields, the shift due to the QED correction to the
moment of the electron would be larger than the mass itself.

Although a fully relativistic treatment may be given
\cite{BanRub2}, we shall, for the proton and neutron quote the
non-relativistic results. For the field ${\bf B}$ along the ${\bf
{z}}$ direction the energy of a proton in the lowest Landau
level and with spin along the magnetic field is
\begin{equation}
E_p(p_z)={\tilde M}_p+{{{p_z}^2}\over {2{\tilde M}_p}}\, ,
\label{nrpe} \end{equation}
where the effective mass ${\tilde M}$ is
\begin{equation}
{\tilde M}_p=M_p-{e\over {2M_p}}\left ({g_p\over 2}-1\right )B\, ;
\label{protmass}
\end{equation}
$M_p$ is the proton's mass and $g_p=5.58$ is the proton's Land\'{e} g
factor.

For the neutron with momentum ${\bf p}$ and spin opposing the
magnetic field, the non-relativistic energy is
\begin{equation}
E({\bf p})=M_n+{e\over {2M_n}}\left ({g_n\over 2}\right )B+
{{{\bf p}^2}\over {2M_n}}\, ,\label{nrne}
\end{equation}
where this time $M_n$ is the neutron's mass and $g_n=-3.82$.

The calculation of the energy of an electron in a strong magnetic
field is more subtle. For fields $B\ge M^2_e/e$ the point formalism
breaks down and we have to solve to one loop QED in the external
field. This problem was treated by Schwinger \cite{Schwinger}.
The energy of an electron with $p_z=0$,
spin up and in the lowest Landau level is \begin{equation}
E_{m,p_z=0}=M_e\left [1+{\alpha\over {2\pi}} \ln \left ( {{2eB}\over
{{M_e}^2}} \right )\right ]\, ;\label{electronener}
\end{equation}
for field strengths of subsequent interest this correction is
negligible.

{}From Eq.~\ (\ref{protmass}) and Eq.~\ (\ref{nrne}) we note that the
neutron becomes stable against $\beta$-decay when the following inequality
is satisfied
\begin{equation}
-{e\over {2M_n}}\left ({g_n\over 2}\right )B
-{e\over {2M_p}}\left ({g_p\over 2}-1\right )B\ge M_n - M_p -
M_e\, ,
\end{equation}
or $B\ge 2\times 10^{14}$ T. On the other hand the proton becomes unstable for
decay into a neutron and a positron whenever
\begin{equation}
\-{e\over {2M_n}}\left ({g_n\over 2}\right )B
-{e\over {2M_p}}\left ({g_p\over 2}-1\right )B\sim 0.12\mu_NB  \ge M_n +
M_e - M_p\, ,\label{threshold}
\end{equation}
or for $B\ge 5\times 10^{14}$ T.

The positron spectrum for the decay of the proton in such a field
is
\begin{equation}
{{d\Gamma}\over {dp_{z,e}}}={4\over 3}{{G_F^2 M_p}\over {(2\pi)^6}}
{{E_e+p_{z,e}}\over {E_e}}(\Delta-E_e)^3\, ;
\end{equation}
where $\Delta= 0.12\mu_NB-M_n+M_p$. For $\Delta\gg M_e$ the total rate
is easily obtained
\begin{equation}
\Gamma={2\over 3}{{G_F^2 M_p}\over {(2\pi)^6}}\Delta^4\, .
\end{equation}
The lifetime is $\tau\sim 1.5\times 10^2 (10^{15}{\rm T}/B)^4$ s.

\section{Screening of Fields by Chiral Symmetry Breaking}
The breaking of the strong interaction chiral symmetry is well
described by the $\sigma$ model. This theory
accounts for the interactions of the Goldstone (more precisely,
the pseudo-Goldstone) modes. We shall show that in very intense
magnetic fields, $B > 1.5\times 10^{14}$ T, the breaking of the
$SU(2)\times SU(2)$ symmetry arranges itself so
that instead of the neutral $\sigma$ field acquiring a vacuum
expectation value it is the charged $\pi$ field that does and at
the same time the magnetic field is screened.

The Hamiltonian density for this problem, including the coupling
of the electromagnetic potential to an external current $\bf j$ is
\begin{eqnarray}
H=&&{1\over 2}\mbox{\boldmath$\nabla$}\sigma\cdot \mbox{\boldmath
$\nabla$}\sigma + {1\over 2}\mbox{\boldmath $\nabla$}\pi_0\cdot
\mbox{\boldmath $\nabla$}\pi_0+ (\mbox{\boldmath $\nabla$}+e{\bf
A}){\pi}^{\dag}\cdot(\mbox{\boldmath $\nabla$}-e{\bf A}){\pi}\nonumber\\
+&&g(\sigma^2+{\bf\pi}\cdot{\bf\pi}-f^2_{\pi})^2+m_{\pi}^2(f_{\pi}-
\sigma) +
{1\over 2}(\mbox{\boldmath $\nabla\times A$})^2 -
 {\bf j}\cdot {\bf A}\, ;
\end{eqnarray}
we have used cylindrical coordinates with \mbox{\boldmath$\rho$}
the two dimensional vector normal to the $z$ direction. In the
limit of large $g$ the radial degree of freedom is frozen out and
the chiral fields may be parameterized by angular variables.
\begin{eqnarray}
\sigma&&=f_{\pi}\cos\chi\, ,\nonumber\\
\pi_0&&=f_{\pi}\sin\chi\cos\theta\, ,\nonumber\\
\pi_x&&=f_{\pi}\sin\chi\sin\theta\cos\phi\, ,\nonumber\\
\pi_y&&=f_{\pi}\sin\chi\sin\theta\sin\phi\, .\label{angvar}
\end{eqnarray}
In terms of which the Hamiltonian density becomes
\begin{eqnarray}
H=&&{{f^2_{\pi}}\over 2}(\mbox{\boldmath$\nabla$}\chi)^2+
{{f^2_{\pi}}\over 2}\sin^2\chi(\mbox{\boldmath$\nabla$}\theta)^2+
{{f^2_{\pi}}\over2}
\sin^2\chi\sin^2\theta(\mbox{\boldmath$\nabla$}\phi- e{\bf A})^2
\nonumber\\
+&&m^2_{\pi}f^2_{\pi}(1-\cos\chi)+
{1\over 2}(\mbox{\boldmath $\nabla\times A$})^2
-{\bf j}\cdot {\bf A}\, .\label{hamdens}
\end{eqnarray}
The angular field $\phi$ can be eliminated by a gauge transformation.

We first look at he case where the external current is due to a thin
wire along the $z$ direction,
\begin{equation}
{\bf j}=I\delta(\mbox{\boldmath $\rho$}){\bf z}\, ;
\end{equation}
An approximate solution, valid for $I/m_\pi >> 1$, is
\begin{eqnarray}
\chi&&=\left\{\begin{array}{ll}
              {\pi\over 2}\ \   & \mbox{for $\rho < \rho_0$}\\
              0     &   \mbox{for $\rho > \rho_0$}\, ,
             \end{array} \right. \nonumber\\
\theta&&={\pi\over 2}\, , \nonumber\\
A&&=\left\{\begin{array}{ll}
     -{I\over 2\pi}\left[ K_0(ef_{\pi}\rho)-
     {{I_0(ef_{\pi}\rho)K_0(ef_{\pi}\rho_0)}/
     I_0(ef_{\pi}\rho_0)}\right]\ \  & \mbox{for $\rho <\rho_0$}\\
   {I\over 2\pi}\ln{\rho\over\rho_0}  & \mbox{for $\rho >\rho_0$}\, ;
    \end{array} \right.
\end{eqnarray}
The parameter $\rho_0$ is determined by minimizing the energy. Doing
this we obtain
\begin{equation}
\rho_0={I\over 2{\sqrt 2}m_\pi f_\pi}\, .\label{rho0}
\end{equation}
We find that the magnetic fields are screened out to a distance
$\rho_0$ from the wire; beyond this distance ordinary
electrodynamics resumes.

Eq.\  (\ref{rho0}) has a very straightforward explanation. It results in
a competition of the magnetic energy density ${1\over 2}B^2$ and the
energy density of the pion mass term $m^2_\pi f^2_\pi (1-\cos\chi)$.
The magnetic field due to the current $I$ is $B={I/ 2\pi\rho}$ and
the transition occurs when $B=B_c$, with $B_c={\sqrt 2}m_\pi
f_\pi\sim 1.5\times 10^{14}$ T. Thus, the chiral fields will adjust
themselves to screen out fields larger than $B_c$ for any current
configuration.

\section{Conclusions}
We have presented several mechanisms that may inhibit the existence of
fields beyond $10^{14}$ T and that may induce unusual hadronic
phenomena as proton $\beta$ decay into a neutron. Which of these
processes will dominate is, for the moment, an open question. As
discussed in the Introduction, fields of such magnitude can only be
realized in connection with conjectured astrophysical phenomena.
Superconducting cosmic strings are supposed to carry currents of
$10^{20}$ A; from Eq.\ (\ref{rho0}) we note that the magnetic field
will be screened out to distances of 40 cm. Should this screening be
incomplete, as perhaps due to penetration by magnetic vortices, then
proton $\beta$ decay could occur. There are suggestions of anomalous
astrophysical positron emission \cite{Cesarsky}.

\nonum
\section{Acknowledgments}
M.\ B.\ was supported in part by the National Science Foundation under
Grant No.\ PHY-89-06641. H.\ R.\ was supported by a SCIENCE EEC
Astroparticles contract.

\end{document}